\documentstyle[12pt]{article}

\advance \topmargin by -\headheight
\advance \topmargin by -\headsep
     
\evensidemargin \oddsidemargin
\begin{document}
\newcommand{\sect}[1]{\setcounter{equation}{0}\section{#1}}
\renewcommand{\theequation}{\thesection.\arabic{equation}}

\topmargin -.6in
\def\lab{\label}
\def\nonu{\nonumber}
\def\rf#1{(\ref{eq:#1})}
\def\lab#1{\label{eq:#1}} 
\def\br{\begin{eqnarray}}
\def\er{\end{eqnarray}}
\def\be{\begin{equation}}
\def\ee{\end{equation}}
\def\0{\nonumber}
\def\lb{\lbrack}
\def\rb{\rbrack}
\def\({\left(}
\def\){\right)}
\def\v{\vert}
\def\bv{\bigm\vert}
\def\lskip{\vskip\baselineskip\vskip-\parskip\noindent}
\relax
\newcommand{\nit}{\noindent}
\newcommand{\ct}[1]{\cite{#1}}
\newcommand{\bi}[1]{\bibitem{#1}}
\def\a{\alpha}
\def\b{\beta}
\def\ca{{\cal A}}
\def\cm{{\cal M}}
\def\cn{{\cal N}}
\def\cf{{\cal F}}
\def\d{\delta} 
\def\D{\Delta}
\def\eps{\epsilon}
\def\g{\gamma}
\def\G{\Gamma}
\def\grad{\nabla}
\def\h{ {1\over 2}  }
\def\hc{\hat{c}}
\def\hd{\hat{d}}
\def\hg{\hat{g}}
\def\hp{ {+{1\over 2}}  }
\def\hm{ {-{1\over 2}}  }
\def\k{\kappa}
\def\l{\lambda}
\def\L{\Lambda}
\def\lg{\langle}
\def\m{\mu}
\def\n{\nu}
\def\o{\over}
\def\om{\omega}
\def\O{\Omega}
\def\p{\phi}
\def\pa{\partial}
\def\pr{\prime}
\def\ra{\rightarrow}
\def\rh{\rho}
\def\rg{\rangle}
\def\s{\sigma}
\def\t{\tau}
\def\th{\theta}
\def\ti{\tilde}
\def\wti{\widetilde}
\def\inte{\int dx }
\def\xb{\bar{x}}
\def\yb{\bar{y}}
\def\tpsi{{\tilde \psi}}
\def\tchi{ {\tilde \chi}}
\def\btpsi{\bar {\tilde \psi}}
\def\btchi{\bar {\tilde \chi}}
\def\tr{\mathop{\rm tr}}
\def\Tr{\mathop{\rm Tr}}
\def\partder#1#2{{\partial #1\over\partial #2}}
\def\ds{{\cal D}_s}
\def\wtwo{{\wti W}_2}
\def\lie{{\cal G}}
\def\alie{{\widehat \lie}}
\def\dlie{{\cal G}^{\ast}}
\def\elie{{\widetilde \lie}}
\def\edlie{{\elie}^{\ast}}
\def\hlie{{\cal H}}
\def\wlie{{\widetilde \lie}}

\def\rlx{\relax\leavevmode}
\def\inbar{\vrule height1.5ex width.4pt depth0pt}
\def\IZ{\rlx\hbox{\sf Z\kern-.4em Z}}
\def\IR{\rlx\hbox{\rm I\kern-.18em R}}
\def\IC{\rlx\hbox{\,$\inbar\kern-.3em{\rm C}$}}
\def\one{\hbox{{1}\kern-.25em\hbox{l}}}

\def\PRL#1#2#3{{\sl Phys. Rev. Lett.} {\bf#1} (#2) #3}
\def\NPB#1#2#3{{\sl Nucl. Phys.} {\bf B#1} (#2) #3}
\def\NPBFS#1#2#3#4{{\sl Nucl. Phys.} {\bf B#2} [FS#1] (#3) #4}
\def\CMP#1#2#3{{\sl Commun. Math. Phys.} {\bf #1} (#2) #3}
\def\PRD#1#2#3{{\sl Phys. Rev.} {\bf D#1} (#2) #3}
\def\PRB#1#2#3{{\sl Phys. Rev.} {\bf B#1} (#2) #3}
\def\PLA#1#2#3{{\sl Phys. Lett.} {\bf #1A} (#2) #3}
\def\PLB#1#2#3{{\sl Phys. Lett.} {\bf #1B} (#2) #3}
\def\JMP#1#2#3{{\sl J. Math. Phys.} {\bf #1} (#2) #3}
\def\PTP#1#2#3{{\sl Prog. Theor. Phys.} {\bf #1} (#2) #3}
\def\SPTP#1#2#3{{\sl Suppl. Prog. Theor. Phys.} {\bf #1} (#2) #3}
\def\AoP#1#2#3{{\sl Ann. of Phys.} {\bf #1} (#2) #3}
\def\PNAS#1#2#3{{\sl Proc. Natl. Acad. Sci. USA} {\bf #1} (#2) #3}
\def\RMP#1#2#3{{\sl Rev. Mod. Phys.} {\bf #1} (#2) #3}
\def\PR#1#2#3{{\sl Phys. Reports} {\bf #1} (#2) #3}
\def\AoM#1#2#3{{\sl Ann. of Math.} {\bf #1} (#2) #3}
\def\UMN#1#2#3{{\sl Usp. Mat. Nauk} {\bf #1} (#2) #3}
\def\FAP#1#2#3{{\sl Funkt. Anal. Prilozheniya} {\bf #1} (#2) #3}
\def\FAaIA#1#2#3{{\sl Functional Analysis and Its Application} {\bf #1} (#2)
#3}
\def\BAMS#1#2#3{{\sl Bull. Am. Math. Soc.} {\bf #1} (#2) #3}
\def\TAMS#1#2#3{{\sl Trans. Am. Math. Soc.} {\bf #1} (#2) #3}
\def\InvM#1#2#3{{\sl Invent. Math.} {\bf #1} (#2) #3}
\def\LMP#1#2#3{{\sl Letters in Math. Phys.} {\bf #1} (#2) #3}
\def\IJMPA#1#2#3{{\sl Int. J. Mod. Phys.} {\bf A#1} (#2) #3}
\def\AdM#1#2#3{{\sl Advances in Math.} {\bf #1} (#2) #3}
\def\RMaP#1#2#3{{\sl Reports on Math. Phys.} {\bf #1} (#2) #3}
\def\IJM#1#2#3{{\sl Ill. J. Math.} {\bf #1} (#2) #3}
\def\APP#1#2#3{{\sl Acta Phys. Polon.} {\bf #1} (#2) #3}
\def\TMP#1#2#3{{\sl Theor. Mat. Phys.} {\bf #1} (#2) #3}
\def\JPA#1#2#3{{\sl J. Physics} {\bf A#1} (#2) #3}
\def\JSM#1#2#3{{\sl J. Soviet Math.} {\bf #1} (#2) #3}
\def\MPLA#1#2#3{{\sl Mod. Phys. Lett.} {\bf A#1} (#2) #3}
\def\JETP#1#2#3{{\sl Sov. Phys. JETP} {\bf #1} (#2) #3}
\def\JETPL#1#2#3{{\sl  Sov. Phys. JETP Lett.} {\bf #1} (#2) #3}
\def\PHSA#1#2#3{{\sl Physica} {\bf A#1} (#2) #3}
\newcommand\twomat[4]{\left(\begin{array}{cc}  
{#1} & {#2} \\ {#3} & {#4} \end{array} \right)}
\newcommand\twocol[2]{\left(\begin{array}{cc}  
{#1} \\ {#2} \end{array} \right)}
\newcommand\twovec[2]{\left(\begin{array}{cc}  
{#1} & {#2} \end{array} \right)}

\newcommand\threemat[9]{\left(\begin{array}{ccc}  
{#1} & {#2} & {#3}\\ {#4} & {#5} & {#6}\\ {#7} & {#8} & {#9} \end{array} \right)}
\newcommand\threecol[3]{\left(\begin{array}{ccc}  
{#1} \\ {#2} \\ {#3}\end{array} \right)}
\newcommand\threevec[3]{\left(\begin{array}{ccc}  
{#1} & {#2} & {#3}\end{array} \right)}

\newcommand\fourcol[4]{\left(\begin{array}{cccc}  
{#1} \\ {#2} \\ {#3} \\ {#4} \end{array} \right)}
\newcommand\fourvec[4]{\left(\begin{array}{cccc}  
{#1} & {#2} & {#3} & {#4} \end{array} \right)}

\begin{titlepage}
\vspace*{-2 cm}
\noindent


\vskip 1 cm
\begin{center}
{\Large\bf  Vertex Operators and Soliton Solutions of  Affine Toda Model with $U(2)$ Symmetry }   \vglue 1  true cm
I. Cabrera-Carnero, { J.F. Gomes}, 
 { G.M. Sotkov} and { A.H. Zimerman}\\

\vspace{1 cm}

{\footnotesize Instituto de F\'\i sica Te\'orica - IFT/UNESP\\
Rua Pamplona 145\\
01405-900, S\~ao Paulo - SP, Brazil}\\

\vspace{1 cm}

\end{center}

\normalsize
\vskip 0.2cm

\begin{center}
{\large {\bf ABSTRACT}}\\
\end{center}
\noindent
The symmetry structure of non-abelian affine Toda model  based on the coset $SL(3)/SL(2)\otimes U(1)$ is studied.  
It is shown 
that the model possess non-abelian Noether symmetry closing   into  a
$q$-deformed $SL(2)\otimes U(1)$ algebra.
Specific  two vertex  soliton solutions  are constructed.
\noindent

\vglue 1 true cm

\end{titlepage}
\section{Introduction}

The family of 2-D relativistic integrable models (IM) known as affine Toda field theories (ATFT) have 
been intensively studied due to their relation
to certain deformations of 2-d conformal field theories (CFT) \cite{fz}, \cite{cm}, \cite{sz}, \cite{sm} 
as well as due to their 
 rich soliton spectrum \cite{holl}, \cite{ot}, \cite{constantinidis}. Certain integrable perturbations 
 of $SU(N)$-WZW models and their gauged
 versions related to ATFT's have also important applications in the condensed matter problems, for 
 example in the description of Heisenberg antiferromagnetic spin chains and ladders \cite{affleck}.
 
  According to
their symmetries we distinguish two classes of ATFT's:  {\it abelian} (A) and {\it non-abelian} (NA) ones.
  The main feature of the non-abelian ATFT's
is that they manifest local or global Noether symmetries (say, $U(1)^{\otimes l}$, $SL(2)\otimes U(1)$, etc ), 
while the  abelian ones do not possess any other symmetries except the discrete $Z_n$ in the case of imaginary   
coupling.  As a consequence the NA-ATFT's admit topological solitons carrying certain Noether charges as well.  Examples of electrically
charged ($U(1)$ or $U(1)\otimes U(1)$) topological solitons have been constructed in refs. \cite{elek}, \cite{dyonic}, \cite{multi},
\cite{local}. As it is well known \cite{DHN}, such finite energy classical solutions play a crucial role in the semiclassical 
quantization as well as for establishing their strong-coupling particle spectra \cite{holl2}. 
 The exact quantum S-matrices of certain T-self dual NA Toda models \cite{annals} with $U(1)$ symmetry have been derived in
ref. \cite{fat}.  

Although the general theory for constructing of Non-Abelian Affine Toda theories  
 is well developed in terms of graded affine algebras (see for instance \cite{laf}, \cite{lez-sav}
 and refs. therein),   its application to  specific NA-ATFT possessing  non-abelian symmetries 
 gives rise  to certain unexpected and interesting structures. 
 One of them is related to the existence of pairs of T-dual NA-ATFT models described in ref. \cite{tdual}. 
  In the present paper we will discuss a $q$-deformed algebraic structure that appears in
    the simplest integrable model of this type, based on the  coset $SL(3)/SL(2)\otimes U(1)$,  constructed in Sect. 2
(see also Sect. 6 of \cite{multi}).  In Sect. 3 we show that  it is  invariant
 under specific non-local and  non-abelian transformations and the corresponding  Noether charges  close
  a $q$-deformed   $SL(2,R)\otimes U(1)$ Poisson bracket algebra.  As a consequence its soliton solutions
 carry both $U(1)$  and isospin charges.  The goal of this paper is to explicitly construct these 1-soliton solutions and to investigate
 their internal non-abelian symmetries.  The two-vertex soliton solutions obtained in Sects. 4 and 5 by the dressing method  
represent a specific subclass of 1-solitons whose spectrum depends on one real parameter only.  Their masses and charges are derived 
in Sect. 5.

\section{Affine Toda Models with Non-Abelian Symmetries}

\vskip 0.5cm

\subsection{The model in the group $G_0$}
As it is well known \cite{lez-sav}, 
the generic NA Toda models  are classified
 according to a $\lie_0 \subset \lie$ embedding  induced
by the grading operator $Q$ decomposing an finite or infinite dimensional Lie algebra 
$\lie = \oplus _{i} \lie _i $ where $
[Q,\lie_{i}]=i\lie_{i}$ and $ [\lie_{i},\lie_{j}]\subset \lie_{i+j}$.  A group
element $g$ can then be written in terms of the Gauss decomposition as 
\be
g= NBM
\label{1}
\ee
where $N=\exp \lie_< $, $B=\exp \lie_{0} $ and
$M=\exp \lie_> $.  The physical fields lie in the zero grade subgroup $B$ 
and the
models we seek correspond to the coset $H_- \backslash G/H_+ $, for $H_{\pm} $ generated by
positive/negative  grade operators.

For consistency with the hamiltonian reduction formalism, the phase space of
the G-invariant WZW model is  reduced by specifying the constant
generators $\eps_{\pm}$ of grade $\pm 1$.  In order to derive 
 an action for $B  $,  invariant under 
\begin{eqnarray}
g\longrightarrow g^{\prime}=\alpha_{-}g\alpha_{+},
\label{2}
\end{eqnarray}
where $\a_{\pm}(z, \bar z)$ lie in the positive/negative grade subgroup
 we have to introduce a set of  auxiliary
gauge fields $A \in \lie _{<} $ and $\bar A \in \lie _{>}$ transforming as 
\begin{eqnarray}
A\longrightarrow A^{\prime}=\alpha_{-}A\alpha_{-}^{-1}
+\alpha_{-}\partial \alpha_{-}^{-1},
\quad \quad 
\bar{A}\longrightarrow \bar{A}^{\prime}=\alpha_{+}^{-1}\bar{A}\alpha_{+}
+\bar{\partial}\alpha_{+}^{-1}\alpha_{+},
\label{3}
\end{eqnarray}
where $\pa  = \pa_t +\pa_x, \bar \pa = \pa_t - \pa_x$.
 The resulting action is the $G/H (= H_- \backslash G/H_+ )$
 gauged WZW    
\begin{eqnarray}
S_{G/H}(g,A,\bar{A})&=&S_{WZW}(g)
\nonumber
\\
&-&\frac{k}{4\pi}\int d^2x Tr\( A(\bar{\partial}gg^{-1}-\epsilon_{+})
+\bar{A}(g^{-1}\partial g-\epsilon_{-})+Ag\bar{A}g^{-1}\) .
\nonumber
\end{eqnarray}
Since the action $S_{G/H}$ is $H$-invariant,
 we may choose $\alpha_{-}=N_{}^{-1}$
and $\alpha_{+}=M_{}^{-1}$. From the orthogonality  of the graded 
subpaces, i.e. $Tr( \lie _i\lie _j ) =0, i+j \neq 0$, we find
\begin{eqnarray}
S_{G/H}(g,A,\bar{A})&=&S_{G/H}(B,A^{\prime},\bar{A}^{\prime})
\nonumber
\\
&=&S_{WZW}(B)-\frac{k}{4\pi}
\int d^2x Tr[-A^{\prime}\epsilon_{+}-\bar{A}^{\prime}\epsilon_{-}
+A^{\prime}B\bar{A}^{\prime}B^{-1}],
\label{14}
\end{eqnarray}
where we have introduced the WZW model action 
\begin{eqnarray}
S_{WZW}=- \frac{k}{4\pi }\int d^2xTr(g^{-1}\partial gg^{-1}\bar{\partial }g)
+\frac{k}{24\pi }\int_{D}\epsilon^{ijk}
Tr(g^{-1}\partial_{i}gg^{-1}\partial_{j}gg^{-1}\partial_{k}g)d^3x.
\label{3a}
\end{eqnarray}
Performing the integration  over the auxiliary fields $A$ and $\bar A$, 
we find  the effective action
\be
S = S_{WZW} (B) - {{k\o {2\pi}}} \int Tr \( \eps_+ B  \eps_- B^{-1}\)d^2x
\label{spm}
\ee
which describes integrable perturbations of the $\lie_0$-WZNW model. 
 Those perturbations are classified in
terms of the possible constant grade $\pm 1$ operators $\eps _{\pm}$.
The equations of motion associated to action  (\ref{spm}) are 
\br
\bar \pa (B^{-1} \pa B ) + [\eps_- ,B^{-1} \eps_+ B] = 0, \quad 
 \pa (\bar \pa B B^{-1}) - [\eps_+ ,B \eps_- B^{-1}] = 0
 \label{ls}
 \er
 For the $\lie = \hat {SL}(3)$ case with homogeneous gradation, $Q = d$ and $\eps_{\pm} = \mu \l_2 \cdot H^{(\pm 1)}$
   and $B= nam$, where
 \br
 n=e^{\tilde \chi_1 E_{-\a_1}}e^{\tilde \chi_2 E_{-\a_2}}e^{\tilde \chi_3 E_{-\a_1 -\a_2}}, \quad
 a = e^{R_1 \l_1 \cdot H + R_2 \l_2 \cdot H}, \quad
 m=e^{\tilde \psi_1 E_{\a_1}}e^{\tilde \psi_2 E_{\a_2}}e^{\tilde \psi_3 E_{\a_1 +\a_2}}
 \label{Bnam}
 \er
 we find the explicit form for the action
 \br
S_{eff} &=& -{{k}\o {8\pi }} \int dz d\bar z  \( {{1}\o {3}} (2  \pa R_1 \bar \pa R_1 - \pa R_1 \bar \pa R_2 
- \pa R_2 \bar \pa R_1 + 2 \pa R_2 \bar \pa R_2 )   \right. \nonu \\
& + & \left. 2\pa \tilde \chi_1 \bar \pa \tilde \psi_1 e^{R_1} 
+2\pa \tilde \chi_2 \bar \pa \tilde \psi_2 e^{R_2} 
+2 (\pa \tilde \chi_3 - \tilde \chi_2 \pa \tilde \chi_1)(\bar \pa \tilde \psi_3 - \tilde \psi_2 \bar \pa \tilde \psi_1)e^{R_1 + R_2} -V \)
\label{6.6np}
\er
 where $V= \mu^2 \(\l_2^2 +\tilde \psi_2 \tilde \chi_2 e^{R_2} + \tilde \psi_3 \tilde \chi_3 e^{R_1+R_2}\)$.

 \subsection{Reduction to the coset $G_0/G_0^0$}

We now introduce the subalgebra 
$\lie_0^0$ such that $[\lie_0^0, \eps_{\pm}] = 0$ as an additional  ingredient which  characterizes 
 the symmetry of action (\ref{spm}) under chiral transformations
\br
B^{\pr} = \bar \Omega (\bar z) B \Omega (z), \quad \bar \Omega , \Omega \in G_0^0
\label{4}
\er
As consequence of symmetry under (\ref{4}), the following chiral conservation laws are 
derived from the equations of motion (\ref{ls}),
\br
\bar \pa Tr(X B^{-1} \pa B ) = 
\pa Tr(X \bar \pa B B^{-1}) = 0, \quad  X \in \lie_0^0
\label{5}
\er
In order to reduce the model to the coset $G_0/G_0^0$, we impose the subsidiary constraints 
\br
 J_{X} =Tr(X B^{-1} \pa B ) = 0, \quad 
 \bar J_{X} = Tr(X \bar \pa B B^{-1}) = 0, \quad  X \in \lie_0^0
\label{6}
\er
which can be incorporated in the action by introducing  the auxiliary gauge fields
 $A^{(0)},\bar A^{(0)} \in \lie_0^0$.  For the models where 
 $\lie_0^0 = U(1)$, \cite{dyonic}, \cite{elek} or $\lie_0^0 = U(1)\otimes U(1)$, \cite{multi},
  the action was constructed  imposing
 symmetry under axial transformations 
 \br
 B^{\pr \pr } = \a_0(\bar z,z) B \a_0(\bar z,z), \quad \a_0 \in G_0^0
 \nonu 
 \er
 and 
 \begin{eqnarray}
 {A^{\pr \pr }}^{(0)} =A^{(0)}-\a_{0}^{-1}\partial \a_{0},
\quad \quad 
 \bar{A ^{\pr \pr}}^{(0)} =\bar{A}^{(0)}
- \a_{0}^{-1}\bar{\partial}\a_{0}.
\label{7}
\end{eqnarray}
For a general non abelian $\lie_0^0$ we can define a second grading structure $Q^{\pr} = \l_1\cdot H$ which decomposes 
 $\lie_0^0$ into positive, zero and negative subspaces, i.e., 
 $\lie_0^0 =  \lie_0^{0,<} \oplus \lie_0^{0,0} \oplus \lie_0^{0,>} $.
 Following the same principle as in  \cite{dyonic}, \cite{elek} and \cite{multi} 
we seek an action  invariant under
\br
 B^{\pr \pr } = \g_0(\bar z,z) \g_-(\bar z,z)B \g_+(\bar z,z)\g_0(\bar z,z), \quad
 \g_0 \in G_0^{0,0},\;  \g_{-} \in G_0^{0,<}, \;  \g_{+} \in G_0^{0,>}
 \nonu 
 \er 
and choose 
 $\g_0(\bar z,z), \g_-(\bar z,z), \g_+(\bar z,z) \in G_0^0$ such that 
$B^{\pr \pr } = \g_0 \g_- B \g_+ \g_0 = g_0^f \in G_0/G_0^0$.  Note that  $B$ 
is  decomposed into the Gauss form according
to the second grading structure $Q^{\pr}$.
Denote $\G_- = \g_0 \g_-$ and $\G_+ = \g_+ \g_0 $.
 Then the  action  
 \br 
S(B,{A}^{(0)},\bar{A}^{(0)} ) &=& S(g_0^f,{A^{\pr}}^{(0)},\bar{A^{\pr}}^{(0)} )  
 = S_{WZW}(B)- 
 {{k\o {2\pi}}} \int Tr \( \eps_+ B  \eps_- B^{-1}\) d^2x\nonu \\   
  &-&{{k\o {2\pi}}}\int Tr\(  A^{(0)}\bar{\partial}B
B^{-1} + \bar{A}^{(0)}B^{-1}\partial B
+ A^{(0)}B\bar{A}^{(0)}B^{-1} + A^{(0)}_0\bar{A}^{(0)}_0 \)d^2x \nonu \\
\label{aa}
\er
is invariant under the transformations $B^{\pr} =  \G_- B \G_+$,
\br
{A^{\pr 0}}_0 &=& A_0^{(0)}  - \g_0^{-1} \pa \g_0, \quad \quad 
{\bar A^{\pr 0}}_0 = \bar A_0^{(0)}  - \g_0^{-1} \bar \pa \g_0, \nonu \\
A^{\pr (0)} &=& \G_- A_{(0)}\G_-^{-1}  - \pa \G_- \G_-^{-1}, \quad \quad 
\bar A^{\pr (0)} = \G_+^{-1}\bar A_{(0)}\G_+  - \G_+^{-1} \bar \pa \G_+ ,
\label{transf}
\er
where 
$A^{(0)} = A^{(0)}_0 + A^{(0)}_{-}$ and $\bar A^{(0)} = \bar A^{(0)}_0 + \bar A^{(0)}_{+}$ and 
 $  A_0^{(0)}, \bar A_0^{(0)} \in \lie_0^{0,0}, A^{(0)}_{-} \in \lie_0^{0,<}, \bar A^{(0)}_{+} \in \lie_0^{0,>} $.

Let us apply the above gauge fixing procedure for the simplest case of $\lie _0 = SL(3,R), \;\; Q=d, \;\; 
\eps_{\pm} = \mu \l_2\cdot H^{(\pm 1)}$  and $\lie_0^0 = SL(2,R)\otimes U(1)$, i.e., for IM defined on the coset 
${\G_-}\setminus {SL(3)}/{\G_+}$ where $\G_{\pm} = exp (\tilde \chi_{\pm} E_{\pm \a_1}) exp ({1\o 2}\l_i \cdot HR_i)$.  
Hence the auxiliary
fields $A_0^{(0)}, \bar A_0^{(0)}, A_-^{(0)}$ and $\bar A_+^{(0)}$ can be parametrized as follows
\br
A_0^{(0)} &=& a_1\l_1 \cdot H + a_2(\l_2 - \l_1 )\cdot H, \quad \bar A_0^{(0)} = \bar a_1\l_1 \cdot H + \bar a_2(\l_2 - \l_1 )\cdot H, \nonu \\
A_-^{(0)} &=& a_{21} E_{-\a_1}^{(0)}, \quad \bar A_+^{(0)} = \bar a_{12} E_{\a_1}^{(0)},
 \label{a0}
\er
where $a_{i}(z, \bar z) , \bar  a_{i}(z, \bar z),  a_{21}(z, \bar z)$ and $  \bar  a_{12}(z, \bar z)$ 
are arbitrary functions and
\br
 g_0^f = e^{\chi_1 E_{-\a_1} + \chi_2 E_{-\a_1-\a_2}}e^{\psi_1 E_{\a_1} + \psi_2 E_{\a_1+\a_2}}.
 \er
 The relation between the fields $\tilde \psi_i,\tilde \chi_i, R_i$ parametrizing the group element (\ref{Bnam}) and the physical fields
of the gauged model  $\psi_1, \chi_1, \psi_2, \chi_2$  parametrizing $g_0^f$ is given by
\br
B = e^{{1\o 2} R_1 \l_1\cdot H + {1\o 2} R_2 \l_2\cdot H} 
e^{\chi_3 E_{-\a_1}}\(g_0^f\)e^{\psi_3 E_{\a_1}}e^{{1\o 2} R_1 \l_1\cdot H + {1\o 2} R_2 \l_2\cdot H}
\label{bg}
\er
or in components, 
\br
 &\tilde \chi_1 = \chi_3 e^{-{1\o 2}R_1},  \quad \quad & \tilde \psi_1 = \psi_3 e^{-{1\o 2}R_1}, \nonu \\
 &\tilde \chi_2 = \chi_2 e^{-{1\o 2}R_2},  \quad \quad & \tilde \psi_2 = \psi_2 e^{-{1\o 2} R_2}, \nonu \\
 &\tilde \chi_3 = \chi_1 e^{-{1\o 2}(R_1 + R_2)},  \quad & \tilde \psi_3 = \psi_1 e^{-{1\o 2}(R_1 + R_2)}. 
 \label{newvaria}
 \er
In order to
calculate the path integral over the auxiliary gauge  fields (\ref{a0}) we first simplify the last term in eqn. (\ref{aa})
\br
     &&  Tr \( A_0^{(0)} \bar A_0^{(0)}  + A^{(0)} g_0^f \bar A^{(0)} g_0^{f -1}  
+ A^{(0)}\bar \pa g_0^f   g_0^{f -1} + \bar A^{(0)}  g_0^{f -1}\pa g_0^f \)\nonu \\
&& =\bar a_i M_{ij}a_j + \bar a_i N_i + \bar N_i  a_i  + \bar a_{12}a_{21}(1+ \psi_2\chi_2) - \bar a_{12}\psi_2\pa \chi_1 -
a_{21} \chi_2 \bar \pa \psi_1 \nonu \\
\label{tr}
\er
where we have introduced the matrix $M$,
\br
M =  \twomat{{4\o 3} + \psi_1 \chi_1}{- {2\o 3}}{-{2\o 3}}{{4\o 3} + \psi_2 \chi_2}, \quad 
\label{matr2}
\er
and the vectors $N$ and $\bar N$,
\br
\bar N = \twovec{-( \bar \pa \psi_1-\bar a_{12}\psi_2 )\chi_1,}{-(\chi_2\bar \pa \psi_2)}, \quad
 N = \twocol{-(  \pa \chi_1- a_{21}\chi_2 )\psi_1}{-(\psi_2\pa \chi_2)}
\label{mat2}
\er
We first evaluate the integral over $a_i$ and $\bar a_i$ in the partition function
\br
Z = \int {\cal D}B{\cal D}A_0^{(0)}{\cal D}\bar A_0^{(0)}{\cal D}A_-^{(0)}{\cal D}\bar A_+^{(0)}e^{-S(B,A^{(0)}, \bar A^{(0)})}
\label{part}
\er
i.e., the Gaussian integral 
\br
\int D\bar {\sl a} D {\sl a}\; \;  e^{\int( \bar {\sl a}M {\sl a} + \bar {\sl a}N +  \bar N {\sl a})} 
= const.\; \; e^{- (\bar N M^{-1}N)}
\nonu
\er
Taking into account the explicit form of $\bar N M^{-1}N$,i.e., 
\br
N M^{-1}N &=& {{4\Delta }\o {3 D}} a_{21} \bar a_{12} + 
2{{\chi_2 a_{21}}\o {3 D}} \( \chi_2 \psi_1 \bar \pa \psi_2 -2 (1+\psi_2 \chi_2) \bar \pa \psi_1\) \nonu \\
&+& 2{{\psi_2 \bar a_{12}}\o {3 D}} \( \psi_2 \chi_1\pa \chi_2 -2(1+ \psi_2 \chi_2)\pa \chi_1\) \nonu \\
&+& {1\o {3D}} \( -({4} + 3\psi_2 \chi_2)\psi_1 \chi_1 \bar \pa \psi_1 \pa \chi_1 
- 2\chi_1 \psi_2 \pa \chi_2 \bar \pa \psi_3 \right. \nonu \\
 &-& \left. 2\chi_2 \psi_1 \bar \pa \psi_2
\pa \chi_1 -  ({4} + 3\psi_1 \chi_1)\psi_2 \chi_2 \bar \pa \psi_2 \pa \chi_2\)\nonu \\
\Delta &=& (1+ \psi_2 \chi_2 )^2 + \psi_1 \chi_1 (1 + {3 \o 4} \psi_2 \chi_2 )\nonu \\
D&=& Det M ={4 \o 3} \( 1 + \psi_1 \chi_1+ \psi_2 \chi_2+ {3 \o 4} \psi_1 \chi_1 \psi_2 \chi_2 \)
\label{delta}
\er
and eqn. (\ref{tr}), we integrate over $a_{21}$ and $\bar a_{12}$.  As a result we derive the  effective action for the gauge fixed
model \footnote{since in this paper we are interested in the solution and symmetries of the classical gauge fixed 
model we are consistently neglecting
all the quantum contributions to $S_{eff}$ comming from the determinant factors,  ghost field action, counter terms, etc }   :
\br
S_{eff} = -{{k}\o {2\pi }} \int dz d\bar z & \( {1\o {\Delta}}( {{\bar \pa \psi_2 \pa \chi_2 }} (1 + 
\psi_1\chi_1 + \psi_2 \chi_2 )  + 
{{\bar \pa \psi_1 \pa \chi_1} }(1 + \psi_2 \chi_2 )  \right. \nonu \\
& \left. - {1\o 2}\( \psi_2 \chi_1 {{\bar \pa \psi_1 \pa \chi_2 }}+ \chi_2 \psi_1 {{\bar \pa \psi_2 \pa \chi_1 }}\)) -V \)
\label{6.6}
\er
where $V =  \mu^2 \({2\o 3} + \psi_1 \chi_1 + \psi_2 \chi_2 \) $.  It appears to be the simplest generalization of the complex sine-Gordon
model \cite{lundregge} and belongs to the same hierarchy of the Fordy-Kulish (multicomponent ) non-linear Schroedinger model \cite{fk},
\cite{jmp}.  One can also derive it as a further hamiltonian reduction of the $A_2^{(1)}$-Homogeneous sine-Gordon model \cite{pousa}.

It is worthwhile to mention that the classical integrability of   the gauged fixed 
 model (\ref{6.6}) is a consequence of the integrability of the corresponding  ungauged model (\ref{6.6np}).  The
 zero curvature (Lax) representation of the IM (\ref{6.6np})(or equivalently  (\ref{ls})) has the well known form:
\br
\pa \bar {{\cal A}} - \bar \pa {\cal A} - [{\cal A},\bar {{\cal A}}] = 0, \quad \quad {\cal A},\bar {{\cal A}} \in \oplus_{i=0, \pm 1}
\lie_i
\label{zzc}
\er
with 
\br
{\cal A} = -B \eps_-B^{-1}, \quad \bar {{\cal A}} =\eps_+ + \bar \pa B B^{-1}
\label{zz}
\er
We next impose the constraints (\ref{6}) on the group element $B$ (\ref{Bnam}), 
i.e. substituting the non-physical fields $R_1, R_2, \tilde
\psi_1$ and $\tilde \chi_1$ by their nonlocal expressions obtained as a solution of the constraints  
(\ref{6}) (see eqns. (\ref{rr}) and  (\ref{psichi}) for
their explicit form).  This gives 
 the Lax connection $ {{\cal A}}, \;   \bar {{\cal A}}$  for the gauged model  (\ref{6.6}).  
   It can be easily verified that substituting (\ref{zz}) into (\ref{zzc}) and 
taking into account (\ref{6}) (or (\ref{rr}) and  (\ref{psichi})),
 one reproduces the equations of motion  derived from the action (\ref{6.6}).  
Then the existence of  an infinite set (of commuting)
conserved charges $P_m, \;\; m=0,1, \cdots$ is a simple consequence of eqn. (\ref{zzc}), namely, 
\br
P_m (t)= Tr \( T(t) \)^m, \quad \pa_t P_m = 0, \quad 
T(t) =\lim \limits_{L \to \infty } {\cal P} exp \int_{-L}^{L} {\cal A}_x(t,x)dx
\nonu
\er
Hence the above described procedure for derivation of the   NA affine Toda model (\ref{6.6}) as gauged $G/H$ two loop WZW models leads to
integrable model by construction.

\section{ Symmetries}

\subsection{Chiral symmetries in the Group $G_0$}

Let us now discuss the  symmetry structure of the ungauged IM on the $SL(3)$ group given by eqn.  (\ref{6.6np}).  
It is generated by the chiral transformation (\ref{4}), 
i.e.,$ B^{\pr} = \bar \Omega (\bar z) B \Omega (z), \quad \bar \Omega , \Omega \in G_0^0 = SL(2) \otimes U(1)$
generated by $\{ \l_1 \cdot H, \l_2 \cdot H,E_{\pm \a_1} \}$.
We make use of the defining representation of $SL(3)$ in terms of 3x3 matrices 
to parametrize  the zero grade group element  $B$ (\ref{Bnam}) in terms of the 
Gauss decomposition, i.e., 
\br B_{1,1} &=& e^{{2\o 3} R_1 + {1\o 3} R_2}, \quad 
B_{1,2} = B_{1,1}\tilde \psi_1, \quad
B_{1,3} =B_{1,1}\tilde \psi_3,  \quad  B_{2,3} =B_{1,1}\(e^{-R_1} \tilde \psi_2 + \tilde \chi_1 \tilde \psi_3\),  \nonu \\
B_{2,1} &=& B_{1,1}\tilde \chi_1, \quad \quad
B_{2,2} = B_{1,1} \( e^{-R_1} +\tilde \psi_1 \tilde \chi_1 \) , \quad \quad 
B_{3,1} = B_{1,1}\tilde \chi_3, \nonu \\
B_{3,2} &=& B_{1,1} \( e^{-R_1}\tilde \chi_2 +\tilde \psi_1 \tilde \chi_3 \) ,  \quad \quad 
B_{3,3} = B_{1,1} \( \tilde \psi_1 \tilde \chi_1 + e^{-R_1}\tilde \psi_2 \tilde \chi_2 + e^{-R_1-R_2} \) , \nonu \\
\label{B33}
\er
We also find for the chiral symmetry transformations  
\br
\Omega_{1,1} &=&e^{{2\o 3} \eps_1 + {1\o 3} \eps_2}, \quad \quad 
\Omega_{1,2} = \Omega_{1,1}\eps_+,\quad \quad 
\Omega_{2,1} = \Omega_{1,1}\eps_-, \nonu \\
\Omega_{2,2} &=& \Omega_{1,1}  \( e^{-\eps_1}  + \eps_- \eps_+ \), \quad \quad 
\Omega_{3,3} = \Omega_{1,1} e^{-\eps_1- \eps_2} , \nonu \\
\Omega_{1,3} &=&\Omega_{2,3} = \Omega_{3,2} = \Omega_{3,1} = 0
\label{o33}
\er
and $\bar \Omega  = \Omega (\eps \rightarrow \bar \eps )$ where $\eps = \eps(z)$ and  $\bar \eps = \bar \eps(\bar z)$.
The infinitesimal chiral transformations for (\ref{4}) yields the following field transformations,
\br
\d R_1 &=& \eps_1 + \bar \eps_1 + 2 \eps_- \tilde \psi_1 + 2 \bar \eps_+ \tilde \chi_1, \nonu \\
\d R_2 &=& \eps_2 + \bar \eps_2 -  \eps_- \tilde \psi_1 -  \bar \eps_+ \tilde \chi_1, \nonu \\
\d \tilde \psi_1 &=&\eps_+ - \eps_1 \tilde \psi_1 + \bar \eps_+ e^{-R_1} - \eps_- \tilde \psi_1^2, \nonu \\
\d \tilde \chi_1 &=&\bar \eps_- - \bar \eps_1 \tilde \chi_1 + \eps_- e^{-R_1} - \bar \eps_+ \tilde \chi_1^2, \nonu \\
\d \tilde \psi_2 &=&\eps_- (\tilde \psi_1 \tilde \psi_2 - \tilde \psi_3) -  \eps_2 \tilde \psi_2, \nonu \\ 
\d \tilde \chi_2 &=& \bar \eps_+ (\tilde \chi_1 \tilde \chi_2 - \tilde \chi_3) - \bar \eps_2 \tilde \chi_2, \nonu \\
\d \tilde \psi_3 &=& -(\eps_1 + \eps_2) \tilde \psi_3 + \bar \eps_+ \tilde \psi_2 e^{-R_1} - \eps_- \tilde \psi_1 \tilde \psi_3, \nonu \\
\d \tilde \chi_3 &=& -(\bar \eps_1 + \bar \eps_2) \tilde \chi_3 +  \eps_- \tilde \chi_2 e^{-R_1} - \bar \eps_+ \tilde \chi_1 \tilde \chi_3
\label{trans33}
\er
As a  consequence of the invariance of the action (\ref{6.6np}) under the chiral transformations 
(\ref{trans33}) we find Noether currents to
correspond to the chiral currents (\ref{5}) associated to the $\lie_0^0$ subalgebra.
For the explicit example of the $\lie_0 = SL(3)$ we have
\br
J_{-\a_1} &=& \pa \tilde \psi_1 - \tilde \psi_1^2 \pa \tilde \chi_1 e^{R_1} + \pa \tilde \chi_2 (\tilde \psi_1 \tilde \psi_2 - \tilde
\psi_3)e^{R_2} \nonu \\
& & + (\pa \tilde \chi_3 - \tilde \chi_2 \pa \tilde \chi_1)(\tilde \psi_1 \tilde \psi_2 - \tilde \psi_3) \tilde \psi_1 e^{R_1 +
R_2} + \tilde \psi_1 \pa R_1 ,\nonu \\
J_{\a_1} &=& \pa \tilde \chi_1 e^{R_1} - \tilde \psi_2 (\pa \tilde \chi_3 - \tilde \chi_2 \pa \tilde \chi_1)e^{R_1+R_2},\nonu \\
J_{\l_1 \cdot H} &=& {1\o 3} (2\pa R_1 + \pa R_2) - \tilde \psi_1 \pa \tilde \chi_1 e^{R_1} + (\tilde \psi_1 \tilde \psi_2 - \tilde \psi_3)
(\pa \tilde \chi_3 - \tilde \chi_2 \pa \tilde \chi_1)e^{R_1+R_2},\nonu \\
J_{\l_2 \cdot H} &=& {1\o 3} (\pa R_1 + 2\pa R_2) - \tilde \psi_2 \pa \tilde \chi_2 e^{R_2} - 
\tilde \psi_3 (\pa \tilde \chi_3 - \tilde \chi_2 \pa \tilde \chi_1)e^{R_1+R_2}\nonu \\
\bar J_{\a_1} &=& \bar \pa \tilde \chi_1 - \tilde \chi_1^2 \bar \pa \tilde \psi_1 e^{R_1} 
+ \bar \pa \tilde \psi_2 (\tilde \chi_1 \tilde \chi_2 - \tilde
\chi_3)e^{R_2} \nonu \\
& &+ (\bar \pa \tilde \psi_3 - \tilde \psi_2 \bar \pa \tilde \psi_1)(\tilde \chi_1 \tilde \chi_2 - \tilde \chi_3) \tilde \chi_1 e^{R_1 +
R_2} + \tilde \chi_1 \pa R_1,\nonu \\
\bar J_{-\a_1} &=& \bar \pa \tilde \psi_1 e^{R_1} - \tilde \chi_2 (\bar \pa \tilde \psi_3 - 
\tilde \psi_2 \bar \pa \tilde \psi_1)e^{R_1+R_2},\nonu \\
\bar J_{\l_1 \cdot H} &=& {1\o 3} (2\bar \pa R_1 + \bar \pa R_2) - \tilde \chi_1 \bar \pa \tilde \psi_1 e^{R_1} 
+ (\tilde \chi_1 \tilde \chi_2 - \tilde \chi_3)
(\bar \pa \tilde \psi_3 - \tilde \psi_2 \bar \pa \tilde \psi_1)e^{R_1+R_2},\nonu \\
\bar J_{\l_2 \cdot H} &=& {1\o 3} (\bar \pa R_1 + 2\bar \pa R_2) - \tilde \chi_2 \bar \pa \tilde \psi_2 e^{R_2} - 
\tilde \chi_3 (\bar \pa \tilde \psi_3 - \tilde \psi_2 \bar \pa \tilde \psi_1)e^{R_1+R_2}
\label{noether}
\er
where $\bar \pa J = \pa \bar J = 0$ and $J = J_{\l_1 \cdot H} h_1 + J_{\l_2 \cdot H} h_2 
+ \sum _{\a = \a_1, \a_2, \a_1+ \a_2} J_{\a} E_{-\a} + J_{-\a} E_{\a}$.  

One can easily verify that the algebra of the local (chiral ) infinitesimal transformations (\ref{trans33}), that leaves invariant the action
of the ungauged IM (\ref{6.6np}) is  $\( SL(2)\otimes U(1)\)_{left} \otimes \( SL(2)\otimes U(1)\)_{right} $.

\subsection{Global symmetries in the coset $G_0 /G_0^0$}

 The reduced  model in the coset  $G_0 /G_0^0$ is obtained by implementing the additional constraints (\ref{6}), i.e. by
 the vanishing of the chiral currents (\ref{6}).  For the $SL(3)$ example this allows the elimination of 
 four degrees of freedom $R_i, \tilde \psi_1$ and $\tilde \chi_1$, i.e., taking into account (\ref{noether}) and (\ref{6}) we find
 \br
 \pa R_1 &=& 2 \tilde \psi_1 \pa \tilde \chi_1 e^{R_1} - \tilde \psi_2 \pa \tilde \chi_2 e^{R_2} + (\pa \tilde \chi_3 - \tilde \chi_2 \pa
 \tilde \chi_1 )(\tilde \psi_3 - 2 \tilde \psi_1 \tilde \psi_2 )e^{R_1 + R_2}, \nonu \\
 \pa R_2 &=& - \tilde \psi_1 \pa \tilde \chi_1 e^{R_1} +2\tilde \psi_2 \pa \tilde \chi_2 e^{R_2} + (\pa \tilde \chi_3 - \tilde \chi_2 \pa
 \tilde \chi_1 )(\tilde \psi_3 + \tilde \psi_1 \tilde \psi_2 )e^{R_1 + R_2}, \nonu \\
 \pa \tilde \psi_1 &=& \tilde \psi_3 \pa \tilde \chi_2 e^{R_2}, \nonu \\
 \pa \tilde \chi_1 &=& \tilde \psi_2 (\pa \tilde \chi_3 - \tilde \chi_2 \pa \tilde \chi_1) e^{R_2}, \nonu \\
 \bar \pa R_1 &=& 2 \tilde \chi_1 \bar \pa \tilde \psi_1 e^{R_1} - \tilde \chi_2 \bar \pa \tilde \psi_2 e^{R_2} + 
 (\bar \pa \tilde \psi_3 - \tilde \psi_2 \bar \pa
 \tilde \psi_1 )(\tilde \chi_3 - 2 \tilde \chi_1 \tilde \chi_2 )e^{R_1 + R_2}, \nonu \\
 \bar \pa R_2 &=& - \tilde \chi_1 \bar \pa \tilde \psi_1 e^{R_1} +2\tilde \chi_2 \bar \pa \tilde \psi_2 e^{R_2} + 
 (\bar \pa \tilde \psi_3 - \tilde \psi_2 \bar \pa
 \tilde \psi_1 )(\tilde \chi_3 + \tilde \chi_1 \tilde \chi_2 )e^{R_1 + R_2}, \nonu \\
 \bar \pa \tilde \psi_1 &=& \tilde \chi_2 (\bar \pa   \tilde \psi_3 - \tilde \psi_2 \bar \pa \tilde \psi_1)e^{R_2}, \nonu \\
 \bar \pa \tilde \chi_1 &=& \tilde \chi_3 \bar \pa \tilde \psi_2  e^{R_2}
 \label{ee}
 \er
  In order to eliminate    the unphysical fields $R_i, \tilde \psi_1$ and $\tilde \chi_1$ we recall 
  eqn. (\ref{newvaria}) relating the fields
  of the gauged and ungauged models (\ref{6.6}) and (\ref{6.6np}) respectively. 
   In terms of these variables the transformations (\ref{trans33}) become,
 \br
 \d \psi_1 &=& {1\o 2} (-\eps_1 -\eps_2 + \bar \eps_1 +\bar \eps_2) \psi_1 - {1\o 2} \eps_- \psi_1 \tilde \psi_1 + 
 \bar \eps_+ \( \psi_2e^{-{1\o 2} R_1} + {1\o 2} \psi_1 \tilde \chi_1\) , \nonu \\
 \d \chi_1 &=& {1\o 2} (\eps_1 +\eps_2 - \bar \eps_1 -\bar \eps_2) \chi_1 - {1\o 2} \bar \eps_+ \chi_1 \tilde \chi_1 +  
 \eps_- \( \chi_2e^{-{1\o 2} R_1} + {1\o 2} \chi_1 \tilde \psi_1\) , \nonu \\
 \d \psi_2 &=& \eps_- \({1\o 2} \psi_2 \tilde \psi_1 - \psi_1 e^{-{1\o 2} R_1} \) - {1\o 2} \bar \eps_+ \psi_2 \tilde \chi_1 + 
 {1\o 2} (-\eps_2 + \bar \eps_2)\psi_2, \nonu \\
 \d \chi_2 &=& \bar \eps_+ \({1\o 2} \chi_2 \tilde \chi_1 - \chi_1 e^{-{1\o 2} R_1} \) - {1\o 2}  \eps_- \chi_2 \tilde \psi_1 +
  {1\o 2} (\eps_2 - \bar \eps_2)\chi_2,
 \label{tranf44}
 \er
 where $\eps_i, \bar \eps_i, \eps_{\pm}$ and $\bar \eps_{\pm}$  satisfy now certain restrictions \footnote{comming from the requirement of
 the invariance of constraints equations (\ref{ee})} (see eqn.
  (\ref{epsr}) below ) which forces them to be constants.
 By simplifying eqn. (\ref{ee}) we obtain the nonlocal fields $R_i$ in the form:
 \br
 \pa R_1 &=& {{\psi_1 \pa \chi_1} \o {\Delta}}( 1+{3\o 2}\psi_2 \chi_2)  - {{\psi_2 \pa \chi_2 }\o {\Delta}}( \Delta_2 + 
 {{3\o 2}} \psi_1 \chi_1), \nonu \\
 \pa R_2 &=& {{\psi_1 \pa \chi_1  }\o {\Delta}} + {{\psi_2 \pa \chi_2 }\o {\Delta}}( 2 \Delta_2 + {{3\o 2}} \psi_1 \chi_1), \nonu \\
\bar  \pa R_1 &=&  
{{\chi_1 \bar \pa \psi_1 } \o {\Delta}} ( 1+{3\o 2}\psi_2 \chi_2)  - {{\chi_2 \bar \pa \psi_2 }\o {\Delta}}( \Delta_2 + 
{{3\o 2}} \psi_1 \chi_1),\nonu \\
 \bar \pa R_2 &=&
{{\chi_1 \bar \pa \psi_1  }\o {\Delta}} + {{\chi_2 \bar \pa \psi_2 }\o {\Delta}}( 2 \Delta_2 + {{3\o 2}} \psi_1 \chi_1)
\label{rr}
\er 
where $\Delta = (1+ \psi_2 \chi_2 )^2 + \psi_1 \chi_1 (1+ {3\o 4} \psi_2 \chi_2 ), \quad \Delta_2 = 1+ \psi_2 \chi_2$.
 In addition we find 
 \br
  \pa \tilde \chi_1 &=&  {{\psi_2}\o {\Delta}}\( \pa \chi_1 \Delta_2 -{1\o 2}\chi_1  \psi_2  \pa \chi_2 \)e^{-{1\o 2}R_1}, \nonu \\
 \pa \tilde \psi_1  &=&  
 {{\psi_1}\o {\Delta}} \( \pa \chi_2 (1+ \psi_1 \chi_1 + \psi_2 \chi_2) - {1\o 2} \chi_2 \psi_1 \pa \chi_1\) e^{-{1\o 2}R_1} , \nonu \\
 \bar \pa \tilde \psi_1  &=& {{\chi_2}\o {\Delta}}\( \bar \pa \psi_1 \Delta_2 -{1\o 2} \psi_1\chi_2  \bar \pa \psi_2\)e^{-{1\o 2}R_1}, \nonu \\ 
\bar \pa \tilde \chi_1  &=& {{\chi_1}\o {\Delta}} \( \bar \pa \psi_2 (1+ \psi_1\chi_1 + \psi_2\chi_2) - 
{1\o 2} \chi_1\psi_2 \bar \pa \psi_1 \)e^{-{1\o 2}R_1} . 
  \label{psichi}
  \er
We next define  the  conserved topological currents
\br
j_{R_i, \mu} = \eps_{\mu  \nu} \pa_{\nu}R_i, \quad i=1,2, \quad 
j_{\tilde \psi_1, \mu} = \eps_{\mu \nu} \pa_{\nu}\tilde \psi_1, \quad
 j_{\tilde \chi_1, \mu} = \eps_{\mu \nu} \pa_{\nu}\tilde \chi_1, 
 \label{noether1}
 \er
 Using the equations of motion derived from (\ref{6.6}), one can confirm 
  the following conservation laws
 \br
\bar \pa j = \pa \bar j, \quad j=  j_{\tilde \psi_1}, \quad j_{\tilde \chi_1}, \quad j = j_{R_i}, i=1,2, 
\label{jbarj}
\er
where $j = {1\o 2}( j_0 +j_1), \quad \bar j = {1\o 2}( j_0 -j_1)$. Notice that (\ref{rr}) and (\ref{psichi}) define 
 the non local fields $R_1, R_2, \tilde \psi_1,\tilde \chi_1$  in terms of the physical 
 fields $\psi_1, \psi_2, \chi_1$ and $\chi_2$.   
 Hence the conservation of the currents defined by the r.h.s. of (\ref{rr}) and (\ref{psichi}) is non-trivial and requires the use of the
 equations of motion.

\subsection{Algebra of the global symmetries}

The simplest way to derive the algebra of symmetries of gauged IM (\ref{6.6})  (generated by transformations (\ref{tranf44}) ) is to realize
the charges of {\it non-chiral} conserved currents (\ref{rr}), (\ref{psichi}) and (\ref{noether1}):
\br
Q_1 &=& {1\o 3}\int (2 \pa_x R_1 + \pa_x R_2)dx, \quad \quad Q_2 = {1\o 3}\int (\pa_x R_1 +2 \pa_x R_2)dx, \nonu \\
Q_{\tilde \chi_1} &=& Q_- = \int \pa_x \tilde \chi_1dx, \quad \quad Q_{\tilde \psi_1} = Q_+ = \int \pa_x \tilde \psi_1dx
\label{cargas}
\er
in terms of the canonical momenta
\br
\Pi_{\psi_1} &=& {{\d {\cal L}}\o {\d \dot \psi_1}} = {{-k}\o {2\pi }}\({{\pa \chi_1}\o {\Delta}} (1 + \psi_2 \chi_2) 
-{1\o 2}{{\pa \chi_2}\o {\Delta}}\chi_1 \psi_2\) , \nonu \\
\Pi_{\psi_2} &=& {{\d {\cal L}}\o {\d \dot \psi_2}} = {{-k}\o {2\pi }}\({{\pa \chi_2}\o {\Delta}} (1 + \psi_1 \chi_1+ \psi_2 \chi_2) 
-{1\o 2}{{\pa \chi_1}\o {\Delta}}\chi_2 \psi_1\), \nonu \\
\Pi_{\chi_1} &=& {{\d {\cal L}}\o {\d \dot \chi_1}} = {{-k}\o {2\pi }}\({{\bar \pa \psi_1}\o {\Delta}} (1 + \psi_2 \chi_2)  
-{1\o 2}{{\bar \pa \psi_2}\o {\Delta}}\chi_2 \psi_1\), \nonu \\
\Pi_{\chi_2} &=& {{\d {\cal L}}\o {\d \dot \chi_2}} = {{-k}\o {2\pi }}\({{\bar \pa \psi_2}\o {\Delta}} (1 + \psi_1 \chi_1+\psi_2 \chi_2)  
-{1\o 2}{{\bar \pa \psi_1}\o {\Delta}}\chi_1 \psi_2\),
\label{mom}
\er
By substituting (\ref{mom}) in eqns. (\ref{rr}) and (\ref{psichi}) we obtain 
\br
\pa R_1 &=& {{-2\pi}\o {k }}\( \psi_1 \Pi_{\psi_1} - \psi_2 \Pi_{\psi_2}\), \quad 
\bar \pa R_1 = {{-2\pi }\o {k}}\(\chi_1 \Pi_{\chi_1} - \chi_2 \Pi_{\chi_2}\), 
\nonu \\
\pa R_2 &=& {{-2\pi }\o {k}}\(\psi_1 \Pi_{\psi_1} +2 \psi_2 \Pi_{\psi_2}\), \quad 
\bar \pa R_2 = {{-2\pi }\o {k}}\(\chi_1 \Pi_{\chi_1} +2 \chi_2 \Pi_{\chi_2}\),\nonu \\
\pa \tilde \chi_1 &=& {{-2\pi }\o {k}}\psi_2 \Pi_{\psi_1}e^{-{1\o 2}R_1}, 
\quad  \quad \bar \pa \tilde \chi_1 = {{-2\pi }\o {k}}\chi_1 \Pi_{\chi_2} e^{-{1\o 2}R_1}\nonu \\
\pa \tilde \psi_1 &=& {{-2\pi }\o {k}}\psi_1 \Pi_{\psi_2}e^{-{1\o 2}R_1},\quad  \quad
\bar \pa \tilde \psi_1 = {{-2\pi }\o {k}}\chi_2 \Pi_{\chi_1} e^{-{1\o 2}R_1}
\label{cpmom}
\er
In order to calculate the field transformations 
\br
\d_{\pm}\psi_i &=& \{Q_{\pm}, \psi_i \} \eps_{\pm}^g, \quad \quad \d_{\pm}\chi_i = \{Q_{\pm}, \chi_i \} \eps_{\pm}^g, \nonu \\
\d_{j}\psi_i &=& \{Q_{j}, \psi_i \} \eps_{j}^g, \quad \quad \d_{j}\chi_i = \{Q_{j}, \chi_i \} \eps_{j}^g
\label{1}
\er
we use the canonical Poisson brackets (PB) (where $\eps^g$ are constant parameters)
\br
\{ \Pi_{\phi_i} (x), \phi_k(y)\} = \d_{ik} \d (x-y), \quad \quad \phi_k = \psi_i, \chi_i
\label{delta}
\er
 and also few consequences of (\ref{delta}) and (\ref{cpmom}):
\br
\{ \pa_x R_1 (x), \psi_i(y)\} &=&  (-1)^{i+1}\psi_i(y)\d (x-y), \quad   \{ \pa_x R_1 (x), \chi_i(y)\} = (-1)^{i} \chi_i(y)\d (x-y),\nonu \\
\{ \pa_x R_2 (x), \psi_2(y)\}  &=& 2\psi_2(y)\d (x-y), \quad  \{ \pa_x R_2 (x), \chi_2(y)\}  =  -2\chi_2(y)\d (x-y),
\label{rcamp}
\er
etc.   Evaluating the corresponding PBs we find the field transformations we seek:
\br
\d_+ \chi_1 &=& {1\o 2} \( \chi_2 e^{-{1\o 2}R_1} + {1\o 2}\chi_1 \tilde \psi_1 \)\eps_+^g , \quad  
  \d_- \chi_1 = {1\o 4}  \chi_1 \tilde \chi_1\eps_-^g, \nonu \\
 \d_+ \psi_1 &=& -{1\o 4}  \psi_1 \tilde \psi_1\eps_+^g, \quad   
  \d_- \psi_1 = -{1\o 2} \( \psi_2 e^{-{1\o 2}R_1} + {1\o 2}\psi_1 \tilde \chi_1 \)\eps_-^g,  \nonu \\
\d_+ \chi_2 &=& -{1\o 4}  \chi_2 \tilde \psi_1\eps_+^g, \quad 
   \d_- \chi_2 = {1\o 2} \( \chi_1 e^{-{1\o 2}R_1} - {1\o 2}\chi_2 \tilde \chi_1 \)\eps_-^g, \nonu \\
 \d_+ \psi_2 &=& -{1\o 2} \( \psi_1 e^{-{1\o 2}R_1} - {1\o 2}\psi_2 \tilde \psi_1 \)\eps_+^g, 
 \quad \d_- \psi_2 = {1\o 4}  \psi_2 \tilde \chi_1\eps_-^g
 \label{delta+}
 \er
Note that the above transformations are nonlocal due to the presence of $\tilde \psi_1$, 
 $\tilde \chi_1$ and $R_1$  which are defined in terms of
integrals of the fields $\psi_i$, $\chi_i$ and their derivatives:
\br
\tilde \psi_1(x) = {1\o 2} \int \eps (x-y) \( \psi_1(y) \Pi_{\psi_2}(y) -  \chi_2(y) \Pi_{\chi_1}(y)\) e^{-{1\o 2}R_1(y)} dy, \nonu \\
R_1 (x) = {1\o 2} \int \eps (x-y) \( \psi_1(y) \Pi_{\psi_1}(y) -  \psi_2(y) \Pi_{\psi_2}(y) - \chi_1(y) \Pi_{\chi_1}(y)+
\chi_2(y) \Pi_{\chi_2}(y)\)dy\nonu \\
\label{int}
 \er
and $\tilde \chi_1 = \tilde \psi_1 (\psi_1 \leftrightarrow \psi_2, \;\; \chi_1 \leftrightarrow \chi_2)$.
   Instead the transformations generated by the charges $Q_1$ and $Q_2$ have the 
following simple, local and linear in the fields form:
 \br
\d_1 \chi_1 &= - \chi_1\eps_1^g, \quad \quad \d_2 \chi_1 =& - \chi_1\eps_2^g, \nonu \\
 \d_1 \psi_1 &=   \psi_1\eps_1^g, \quad \quad \d_2 \psi_1 =&  \psi_1\eps_2^g, \nonu \\
\d_1 \chi_2 &= 0, \quad \quad \d_2 \chi_2 =& -  \chi_2\eps_2^g, \nonu \\
 \d_1 \psi_2 &= 0, \quad \quad \d_2 \psi_2 =&   \psi_2\eps_2^g 
 \label{deltar}
 \er
Observe that  the above transformations coincide precisely with the transformations (\ref{tranf44}) derived in Sect. 3.2 provided
the following identities take place,  
\br
2\eps_1^g &=& \bar \eps_1 - \eps_1, \quad 2\eps_2^g = \bar \eps_2 - \eps_2,\nonu \\
{1\o 2}\eps_+^g &=& \eps_-, \quad -{1\o 2}\eps_-^g =\bar \eps_+
\label{epsr}
\er

The PB algebra  of the charges $Q_{\pm}$, $Q_1, Q_2$ can be evaluated with help of eqns. (\ref{delta}),  (\ref{rcamp}) and  (\ref{int})
yielding the following deformed structure
\br
 \{ Q_1, Q_{\pm}\} &=&\pm  Q_{\pm}, \nonu \\
 \{ Q_2, Q_{\pm}\} &=& 0, \nonu \\
 \{ Q_+, Q_{-} \} &=&- ({{2\pi}\o {k}})^{2}\int \pa_x e^{- R_1} dx = 2\kappa  ({{2\pi}\o {k}})^{2}sinh \(  Q_1  -{1\o 2}Q_2\) 
 \label{algcharg}
 \er
where $\kappa = exp ({{1}\o 2} (R_1(\infty) + R_1 (-\infty)))$.
 Note that $\kappa$ is a constant operator commuting with all the other
generators.  Finally one can verify the invariance of the gauged IM (\ref{6.6}) under the above nonlocal transformations by calculating the
PB's of the charges $Q_{\pm}, Q_i$ with the hamiltonian of the model
\br
{ H} &=& \int dx ((1+ \psi_2 \chi_2 ) \Pi_{\chi_2}\Pi_{\psi_2} + {1\o 2} \psi_2 \chi_1 \Pi_{\chi_1}\Pi_{\psi_2}
+ (1+ \psi_1 \chi_1 +\psi_2 \chi_2 )\Pi_{\chi_1}\Pi_{\psi_1} \nonu \\
 &+& {1\o 2} \psi_1 \chi_2 \Pi_{\chi_2}\Pi_{\psi_1}
 + \psi_1 ^{\pr} \Pi_{\psi_1}+ \psi_2 ^{\pr} \Pi_{\psi_2}
-\chi_1 ^{\pr} \Pi_{\chi_1} - \chi_2 ^{\pr} \Pi_{\chi_2} +V)
\label{ham}
\er
After a tedious but straightforward calculations we find that 
\br
\{ Q_{\pm}, H\} = 0, \quad \quad \{ Q_{i}, H\} = 0
\nonu
\er
Hence the IM in consideration is invariant under the algebra (\ref{algcharg}), which after certain rescaling of the generators (see for
example ref. \cite{leclair}) can be
identified with the $q$-deformed $SL(2, R)\otimes U(1)$ PB algebra.

\section{Dressing Transformations and Vertex Operators}

As it is well known \cite{ot}, \cite{bb} the dressing transformation and the vertex operators method represents a
powerfull tool for the construction of solitons solutions for the affine Toda models.
Let us consider two arbitrary solutions $B_s \in \hat G_0, \; s=1,2 $ of eqns. (\ref{ls}) 
 written for the case of $A_2^{(1)}$
 extended by $d$ and the central term $c$, i.e.
 \br
 B_s = g_{0s} e^{\nu_s c + \eta_s d}, 
 \nonu
 \er
 The corresponding Lax (L-S) connections (\ref{zz}) ${\cal{A}}(s) = {\cal{A}}(B_s), \; {\bar {\cal{A}}}(s) = 
 {\bar {\cal{A}}}(B_s) $ are related by gauge (dressing) transformations $\theta_{-,+} = \exp {\lie_{<,>}}$,
 \br
 {\cal{A}}_{\mu}(2)= \theta_{\pm} {\cal{A}}_{\mu}(1)\theta_{\pm}^{-1} + \( \pa_{\mu}\theta_{\pm}\) \theta_{\pm}^{-1}
 \label{4.1}
 \er
 They leave invariant the equations of motion (\ref{ls}) as well as the auxiliar linear problem, i.e. the pure gauge 
${\cal{A}}_{\mu}$ defined in terms of the monodromy matrix $T(B_s)$,
\br
\(\pa_{\mu} - {{\cal{A}}(B_s)}_{\mu}\)T_s (B_s) = 0
\label{4.2}
\er
The consistency of equations (\ref{4.1}) and (\ref{4.2}) imply the  following relations 
\br
T_2 = \theta_{\pm} T_1,\; \; i.e. \; \; \theta_+ T_1 = \theta_- T_1 g^{(1)}
\label{4.3}
\er
where $g^{(1)} \in \hat {G}$ is an arbitrary constant element of the corresponding affine group.  Suppose $T_1 =
T_0(B_{vac})$ is the vacuum solution, 
\br
B_{vac} \eps_- B_{vac}^{-1} &=& \eps_-, \quad \bar \pa B_{vac}  B_{vac}^{-1}= \mu^2 z c, \nonu \\
{\cal {A}}_{vac} &=&-\eps_-, \quad \bar {{\cal {A}}}_{vac} =\eps_+ +\mu^2z c,
\label{4.4}
\er
and $T_0 = \exp (-z \eps_-) \exp (\bar z \eps_+)$ as one can easily check by using the fact 
that $[ \eps_+, \eps_-] = \mu^2
c$.  According to eqns. (\ref{4.1}) and (\ref{4.3}), every solution $T_2 = T(B)$ can be obtained 
from the vacuum configuration
(\ref{4.4}) by an appropriate gauge transformation $\theta_{\pm}$.  In fact, eqns. (\ref{4.1}) with
${\cal {A}}_{vac}$ and $\bar {{\cal {A}}}_{vac}$ as in eqn. (\ref{4.4}) and 
\br
{\cal {A}}(B) = -B \eps_- B^{-1}, \quad \bar {{\cal {A}}}(B)= \eps_+ + \bar \pa B  B^{-1}
\nonu 
\er
allows to derive $\theta_{\pm}$ as functionals of $B$, i.e. $\theta_{\pm} = \theta_{\pm} (B)$.  We next apply eqns.
(\ref{4.3}), 
\br
\theta_-^{-1} \theta_+ = T_{vac} g^{(1)} T_{vac}^{-1}
\label{4.5}
\er
in order to obtain a non trivial field configuration  $B$  in terms of $g^{(1)}\in G$ and certain highest weight
(h.w.) representation  of the  algebra $A_2^{(1)}$ as we shall see below.  The first step consists in
substituting ${\cal {A}}_{vac},   \bar {{\cal {A}}}_{vac}$ and 
${\cal {A}}(B),   \bar {{\cal {A}}}(B)$ in eqn. (\ref{4.1}) and then solving it grade by grade remembering that
$\theta_{\pm}$ may be decomposed in  the form of infinite products
\br
\theta_- = e^{t(0)}e^{t(-1)}\cdots , \quad \quad \theta_+ =  e^{v(0)}e^{v(1)}\cdots
\nonu 
\er
where $t(-k)$ and $v(k), k=1,2, \cdots $ denote  linear combinations of grade $p=\mp k$ generators. 
 For grade zero we find 
 \br
 t(0) = H(\bar z), \quad \quad e^{v(0)} = B e^{G(z)-\mu^2 z \bar z c} \nonu 
 \er
 where the arbitrary functions $H(\bar z), G(z) \in \lie_0^0$ and are fixed to zero due to the subsidiary constraints
 (\ref{rr}), (\ref{psichi}), i.e., $H(\bar z) = G(z)=0$.  The equations for $v(1), t(-1)$  
appears to be of the form
\br
B^{-1} \pa B - \mu^2 \bar z c = [ v(1), \eps_-], \quad \bar \pa B B^{-1} = [t(-1), \eps_+]+ \mu^2 z c
\nonu 
\er
The next step is to consider certain matrix elements (taken for the h.w. representation $|\l_l >$) of eqn. (\ref{4.5}). 
Since $v(i)|\l_l > = 0$ and $<\l_l | t(-i) =0, i>0$., we conclude that 
\br
<\l_l | B |\l_l >e^{-\m^2z \bar z} = <\l_l | T_0 g^{(1)} T_0^{-1} |\l_l >
\label{4.6}
\er
Taking into account the explicit parametrization of the zero grade subgroup element $B = nam $ (\ref{Bnam}) in terms of the
 fields, $\nu, R_i, \psi_a, \chi_a$ and choosing specific matrix elements we derive their
explicit  space-time dependence, 
\br
\tau_0 \equiv e^{\nu -  \mu^2 z \bar z} &=& < \lambda_0|T_{0}g^{(1)}T_{0}^{-1}|\lambda_0 >,
\nonu \\
\tau_{1} \equiv e^{{1\o 3}(2R_1+R_2) + \nu -  \mu^2 z \bar z}  &=& < \lambda_1|T_{0}g^{(1)}T_{0}^{-1}|\lambda_1 >,
\nonu \\
\tau_{2} \equiv e^{{1\o 3}(R_1+2R_2)  +\nu -  \mu^2 z \bar z} &=& <\lambda_{2}|T_{0}g^{(1)}T_{0}^{-1}|\lambda_{2} >,
\nonu \\
\tau_{\psi_{3}}  \equiv   e^{{1\o 3}(2R_1+R_2) + \nu -  \mu^2 z \bar z} 
\tilde \psi_{3} &=&
 < \lambda_{1}|T_{0}g^{(1)}T_{0}^{-1}E_{-\a_{1}-\a_2}^{(0)}|\lambda_{1} >,
\nonu \\
\tau_{\chi_{3}} \equiv   e^{{1\o 3}(2R_1+R_2) + \nu -  \mu^2 z \bar z} 
\tilde \chi_{3} &=&
 < \lambda_{1}|E_{\a_{1}+\a_2}^{(0)}T_{0}g^{(1)}T_{0}^{-1}|\lambda_{1} >,
\nonu \\
\tau_{\psi_{2}}  \equiv   e^{{1\o 3}(R_1+2R_2) + \nu -  \mu^2 z \bar z} 
\tilde \psi_{2} &=&
 < \lambda_{2}|T_{0}g^{(1)}T_{0}^{-1}E_{-\a_2}^{(0)}|\lambda_{2} >,
\nonu \\
\tau_{\chi_{2}} \equiv   e^{{1\o 3}(R_1+2R_2) + \nu -  \mu^2 z \bar z} 
\tilde \chi_{2} &=&
 < \lambda_{2}|E_{\a_2}^{(0)}T_{0}g^{(1)}T_{0}^{-1}|\lambda_{2} >,
\nonu \\
\tau_{\psi_{1}}  \equiv   e^{{1\o 3}(2R_1+R_2) + \nu -  \mu^2 z \bar z} 
\tilde \psi_{1} &=&
 < \lambda_{1}|T_{0}g^{(1)}T_{0}^{-1}E_{-\a_{1}}^{(0)}|\lambda_{1} >,
\nonu \\
\tau_{\chi_{1}} \equiv   e^{{1\o 3}(2R_1+R_2) + \nu -  \mu^2 z \bar z} 
\tilde \chi_{1} &=&
 < \lambda_{1}|E_{\a_{1}}^{(0)}T_{0}g^{(1)}T_{0}^{-1}|\lambda_{1} >,
\label{4.7}
\er
In order to make the construction of the
solution (\ref{4.7})  complete it remains to specify the constant affine 
group element $g^{(1)}$, which  encodes
the information (including topological properties) about the N-soliton structure 
of eqns. (\ref{ls}). 

Since $\eps_{\pm}$  form a
Heisenberg subalgebra, 
$[\eps_+, \eps_- ]= \mu^2 c$
and we have to calculate the matrix elements ($\tau$-functions ), say
\br
< \l_0 | e^{-z \eps_-} e^{\bar z \eps_+} g^{(1)} e^{-\bar z \eps_+}e^{z \eps_-} | \l_0 >,
\nonu 
\er
it is instructive to introduce the eigenvectors  ($F(\g )$) of $\eps_{\pm}$, i.e.,
\begin{equation}
\lbrack \eps ^{\pm },F(\gamma )]=f^{\pm }(\gamma )F(\gamma ), \; \;\;
\label{fgamma}
\end{equation}
Following ref. \cite{aratyn} (see also \cite{multi}) we find  four non trivial   types of eigenvectors 
\br  
F_{\pm }(\g ) &=& \sum_{n\in Z} \( E_{\pm \a_2}^{(n)} + E_{\pm (\a_1+\a_2)}^{(n)}\) \g^{-n}, \nonu \\
\tilde F_{\pm }(\g ) &=& \sum_{n\in Z} \( E_{\pm \a_2}^{(n)} - E_{\pm (\a_1+\a_2)}^{(n)}\) \g^{-n}
\label{4.12}   
\er
as well as the trivial eigenvectors 
\br
F^0_{\pm} (\g ) = \sum_{n\in Z}E_{\pm \a_1}^{(n)} \g^{-n}
\er
Their eigenvalues are given by 
\br
\lbrack \eps _{+ },F_{\pm }(\gamma )] & =& \pm \mu \gamma F_{\pm }(\gamma ), \quad 
\lbrack \eps _{+ },\tilde F_{\pm }(\gamma )]  = \pm \mu \gamma \tilde F_{\pm }(\gamma ), \quad 
\lbrack \eps _{+ },F_{\pm }^0(\gamma )]  = 0,
\nonu \\
\lbrack \eps _{- },F_{\pm }(\gamma )] & =& \pm \mu \gamma^{-1} F_{\pm }(\gamma ), \quad 
\lbrack \eps _{- },\tilde F_{\pm }(\gamma )]  =\pm \mu \gamma^{-1} \tilde F_{\pm }(\gamma ), \quad 
\lbrack \eps _{- },F_{\pm }^0(\gamma )]  = 0,
\label{4.14}
\er
Notice that $F_{\pm}(\g ), \tilde F_{\pm}(\g )$ and $F_{\pm}^0(\g )$ together with 
\br
\l_i\cdot H(\g ) = \sum_{n\in Z} \l_i\cdot H^{(n)} \g^{-n}, \quad i=1,2
\er
form a new basis for the affine $A_2^{(1)}$ algebra. 
In this basis, we define the affine group element $g^{(1)}$ as 
\br
g^{(1)}= \prod_{a}e^{d_aF_a(\g )}, \quad F_a(\g ) =\{ F_{\pm }, \tilde F_{\pm}, \l_i\cdot H{(\g )}, \;\; i=1,2,\;\; F_{\pm}^0(\g )\}
\nonu 
\er
with the property
\br
T_0g^{(1)}T_0^{-1} &=& \exp (\sum_{a}d_a \rho_a(\g ) F_a(\g ) ) = \prod_{a} \(1 + d_a \rho_a(\g ) F_a(\g )\), \nonu \\ 
\rho_a (\g ) &=& \exp (-z f_a^-(\g ) + \bar z f_a^+(\g ))
\label{tt}
\er
The use of this basis  drastically simplify the calculation of the $\tau $-functions (\ref{4.7}).  
Notice that in the above formula each $F_a^2=0$,  but the
mixed terms  $F_a F_b$ do contribute \cite{aratyn}.

\section{Two-Vertex  Soliton Solutions}
An important question concerns the specific choice of the form  of $g^{(1)}$ that leads to different species of
neutral and charged solitons and breathers.  As in the case of complex sine-Gordon model the 1-soliton 
solutions can be constructed in terms of two
vertex operators, i.e., $g^{(1)}$ chosen in one of the  following forms:
\br
g^{(1)} (\g_1, \g_2)  = e^{d_1 F_+(\g_1)} e^{  d_2 F_-(\g_2)}
\label{4.18a}
\er
\br
\tilde g^{(1)} (\g_1, \g_2)  = e^{\tilde d_1 \tilde F_+(\g_1)}e^{ \tilde  d_2 \tilde F_-(\g_2)}
\label{4.18b}
\er
\br
 g^{(1)}_{01} (\g_1, \g_2)  = e^{ d_{01}  F_+(\g_1)}e^{ \tilde  d_{01} \tilde F_-(\g_2)}
\label{4.18c}
\er
\br
 g^{(1)}_{02} (\g_1, \g_2)  = e^{ \tilde d_{02}  \tilde F_+(\g_1)}e^{   d_{02}  F_-(\g_2)}
\label{4.18d}
\er
For the case given by eqn. (\ref{4.18a}) according to (\ref{tt}) we find for the $\tau$ -functions (\ref{4.7}),
\br 
\tau_0 &=& 1+ 2 d_1d_2 \rho_1(\g_1) \rho_2(\g_2) {{\g_1 \g_2 }\o {(\g_1 - \g_2 )^2}}, \nonu \\
\tau_1 &=& 1+  d_1d_2 \rho_1(\g_1) \rho_2(\g_2) {{\g_1 (\g_1 +\g_2 )}\o {(\g_1 - \g_2 )^2}}, \nonu \\
\tau_2 &=& 1+ 2 d_1d_2 \rho_1(\g_1) \rho_2(\g_2) {{\g_1^2 }\o {(\g_1 - \g_2 )^2}}, \nonu \\
\tau_{\psi_3} &=& \tau_{\psi_2} = d_1 \rho_1(\g_1), \quad \tau_{\chi_3} = \tau_{\chi_2} =d_2 \rho_2(\g_2), \nonu \\
\tau_{\psi_1} &=&  \tau_{\chi_1} = d_1 d_2 \rho_1(\g_1)\rho_2(\g_2) {{\g_1}\o {\g_1 - \g_2}}, 
\label{tauu}
\er
where $\rho_1(\g_1)= \exp (-{{z}\o {\g_1}} + \bar z \g_1)$ and $\rho_2(\g_2)= \exp ({{z}\o {\g_2}} - \bar z \g_2)$.
In order to ensure the reality (and positivity) of the energy of the solution, we require 
that the product $\rho_1(\g_1 ) \rho_2(\g_2 )$ to be real.  This leads to the following parametrization for $\g_i$:
\br
\g_1 = -e^{b-i\a}, \quad \g_2 = e^{b+i\a}, \quad b, \a \in R
\nonu
\er
and for $\rho_i(\g_i )$ we obtain,
\br
\rho_1 = e^{F+iG}, \quad \rho_2 = e^{F-iG}, \quad z = {1\o 2} (x+t), \quad  \bar z = {1\o 2} (t-x)\nonu \\
F= \mu \cos (\a )[-t sh (b) + x ch (b)], \quad G= \mu \sin (\a )[t ch  (b) - x sh (b)]
\nonu
\er
It is convenient to choose the arbitrary complex constants $d_1$ and $d_2$ in the form:
\br
d_1 = {{\g_1- \g_2}\o {\sqrt {2\g_1 \g_2} }}e^{i\theta - \mu Y \cos (\a )ch (b)}, \quad 
d_2 = {{\g_1- \g_2}\o {\sqrt {2\g_1 \g_2} }}e^{-i\theta - \mu Y \cos (\a )ch (b)}
\nonu
\er
where $\theta$ and $Y$ are new arbitrary real constants.  Then the 1-soliton solutions corresponding 
to two vertex $g^{(1)}$ (\ref{4.18a}) takes the following simple form:
\br
e^{\nu - \mu^2 z \bar z} &=& 1+ e^{2\tilde F}, \nonu \\
e^{{{1 }\o 3}(2R_1 + R_2)}&=& {{e^{-\tilde F} + {{1\o 2}(1-\Gamma_1)e^{\tilde F} }}\o {e^{\tilde F}+e^{-\tilde F}}}, \quad \Gamma_1 =
e^{-2i\a}, \nonu \\
e^{{{1 }\o 3}(R_1 + 2R_2)}&=& {{{e^{-\tilde F} -\Gamma_1e^{\tilde F} }}\o {e^{\tilde F}+e^{-\tilde F}}}, \quad \Gamma_2 =
d_1e^{-i\theta+\mu Y \cos (\a )ch (b)} = {{\g_1- \g_2}\o {\sqrt {2\g_1 \g_2} }}, \nonu \\
\psi_1 &=&{{\Gamma_2 e^{i(G+\theta )}}\o {  (e^{\tilde F}+e^{-\tilde F})}}
\( {{e^{-\tilde F}- \Gamma_1e^{\tilde F}}\o {{e^{-\tilde F} + {{1\o 2}(1-\Gamma_1)e^{\tilde F} }}}}\)^{{1\o 2}}, \nonu \\
\chi_1 &=&{{\Gamma_2 e^{-i(G+\theta )}}\o {  (e^{\tilde F}+e^{-\tilde F})}}
\( {{e^{-\tilde F}- \Gamma_1e^{\tilde F}}\o {{e^{-\tilde F} + {{1\o 2}(1-\Gamma_1)e^{\tilde F} }}}}\)^{{1\o 2}}, \nonu \\
\psi_2 &=&{{\Gamma_2 e^{i(G+\theta )}}\o {  (e^{\tilde F}+e^{-\tilde F})}}
\( {{e^{-\tilde F}+e^{\tilde F}}\o {{e^{-\tilde F} + {{1\o 2}(1-\Gamma_1)e^{\tilde F} }}}}\)^{{1\o 2}}, \nonu \\
\chi_2 &=&{{\Gamma_2 e^{-i(G+\theta )}}\o {  (e^{\tilde F}+e^{-\tilde F})}}
\( {{e^{-\tilde F}+e^{\tilde F}}\o {{e^{-\tilde F} + {{1\o 2}(1-\Gamma_1)e^{\tilde F} }}}}\)^{{1\o 2}}
\label{solution}
\er
where $\tilde F(t,x) = F(t, x-Y)$.  The nonlocal fields $\tilde \psi_1$ and $\tilde \chi_1$  (whose  asymptotics 
are important for determining the charges $Q_{\pm}$) are given by:
\br
 \tilde \psi_1 = \tilde \chi_1 = 
 -{{(1+\Gamma_1)e^{2\tilde F}}\o {2 \( 1+ {1\o 2}(1-\Gamma_1) e^{2\tilde F}\)}}
\label{nonloc}
\er
 As it is well known \cite{ot}, \cite{constantinidis}, \cite{dyonic} the energy of such 
 solution is related to the asymptotics of $ln \tau_0$, i.e.,
 \br
 M= E(b=0) = \int _{-\infty}^{\infty} dx T_{00} = 
 -{{2}\o {\b^2}} \int _{-\infty}^{\infty} dx \pa_x ln \tau_0 = {{4\mu }\o {\b^2}}\cos (\a), \quad \b^2 = {{2\pi}\o k}
 \nonu
 \er
in the rest frame $b=0$.  The corresponding Noether charges $Q_1$ and $Q_2$ (see eqns. (\ref{cargas})) are 
defined by the asymptotics of the nonlocal fields $R_i$
\br
Q_1 = -i \(\a +\pi +i ln (\cos (\a))\), \quad 
Q_2 = -i \(\a +\pi +i ln (2\cos (\a))\)
\label{char}
\er
Similarly, for the charges $Q_{\pm}$ we find
\br
Q_+ = Q_- = \tilde \psi_1(\infty ) - \tilde \psi_1(-\infty ) = {{\Gamma_1 + 1}\o {\Gamma_1 -1}} =  icotg (\a )
\nonu
\er
Therefore the spectrum of the above 1-soliton solutions, i.e., $M, Q_1, Q_2$ and $Q_{\pm}$ is determined by the real constant $\a$ only.

The case when $g^{(1)}$ is taken in the form (\ref{4.18b}) is quite similar  to the considered above.  The only difference is that now we
have
\br
\tau_{\psi_3} = - \tau_{\psi_2}, \quad \tau_{\chi_3} = - \tau_{\chi_2}, \quad \tau_{\psi_1} = 
 \tau_{\chi_1}= -\tilde d_1 \tilde d_2 \rho_1 \rho_2 {{\g_1}\o {\g_1 - \g_2}} 
\nonu
\er
and all the other $tau$-functions remain unchanged.  
As a consequence, the mass $\tilde M$ and charges $\tilde Q_1, \tilde Q_2$ are the same
and  $\tilde Q_{\pm }= -Q_{\pm }$.
The cases (\ref{4.18c}) and (\ref{4.18d}) are quite different.  We find that
\br
\tau_0 &=& \tau_2 =1, \quad \tau_1 =   1- d_1\tilde d_2 \rho_1(\g_1) \rho_2(\g_2) {{\g_1}\o {\g_1 - \g_2}}, \nonu \\
\tau_{\psi_1} &=& -  \tau_{\chi_1} = d_1\tilde d_2  \rho_1(\g_1)\rho_2(\g_2) {{\g_1}\o {\g_1 - \g_2}}, \nonu \\
\tau_{\psi_3} &=& -  \tau_{\psi_2} = d_1 \rho_1(\g_1), \quad  \tau_{\chi_3} = -  \tau_{\chi_2} = -\tilde d_2 \rho_2(\g_2)
\nonu
\er
Such solution has vanishing mass and charge $Q_2 =0$.

\section{Concluding Remarks}

Our analysis of the symmetries and 1-soliton solutions of the IM (\ref{6.6}) leaves few interesting open problems:
\begin{itemize}
\item How to construct more general 1-soliton solutions whose spectrum, $M,Q_1, Q_2$ and $Q_{\pm}$ is parametrized by four real parameters
instead of one  $\a$ as in eqn. (\ref{solution})
\item What are the symmetry properties of the 1-solitons (\ref{solution}), 
i.e. to recognize the representations of the $q$-deformed  algebra
(\ref{algcharg}) to which these solitons belong to.
\item About the topological stability of these solitons and of the related  strong coupling particles of the IM (\ref{6.6}) 
\end{itemize}

The complete structure of the solitons (and particles) spectrum of such IM ideed require to answer the above questions.

{\bf Acknowledgments}  One of us (ICC) would like to thank P. Teotonio Sobrinho for discussions. 
 We thank CNPq and Fapesp for support.


\end{document}